\documentclass[runningheads]{llncs}
\usepackage[T1]{fontenc}
\usepackage{graphicx}
\usepackage{url}
\usepackage{booktabs}
\usepackage{hyperref}
\usepackage{footmisc}
\begin{document}
\title{Toward Greener Background Processes: Measuring Energy Cost of Autosave Feature}

\titlerunning{Toward Greener Background Processes}
\author{Maria Küüsvek\inst{1} \and
Hina Anwar\inst{1}\orcidID{0000-0002-4725-4636} }
\authorrunning{M. Küüsvek and H. Anwar}
\institute{University of Tartu, Institute of Computer Science, Tartu, Estonia \\
\email{\{maria.kuusvek,hina.anwar\}@ut.ee}}
\maketitle              % typeset the header of the contribution
\begin{abstract}
Background processes in desktop applications are often overlooked in energy consumption studies, yet they represent continuous, automated workloads with significant cumulative impact. This paper introduces a reusable process for evaluating the energy behavior of such features at the level of operational design. The process works in three phases: 1) decomposing background functionality into core operations, 2) operational isolation, and 3) controlled measurements enabling comparative profiling. We instantiate the process in a case study of autosave implementations across three open-source Python-based text editors. Using 900 empirical software-based energy measurements, we identify key design factors affecting energy use, including save frequency, buffering strategy, and auxiliary logic such as change detection. We give four actionable recommendations for greener implementations of autosave features in Python to support sustainable software practices.

\end{abstract}

\section{Introduction}

Software energy consumption is emerging as a critical concern in sustainable computing, particularly in battery-constrained devices. While hardware efficiency has advanced, the energy footprint of software, especially background features, remains under-optimized. Huawei estimates that ICT will consume 9\% of global energy by 2025, with projections reaching 21\% by 2030 \cite{andrae2017total,jones2018stop}. Yet, only 18\% of developers consider energy efficiency during development \cite{pang2015programmers}.

Background process can be defined as a program component that executes alongside the primary foreground workflow, starting automatically or indirectly, running with no UI, and continuing to run despite user interactions\footnote{\url{https://en.wikipedia.org/wiki/Background\_process}}
\footnote{\url{https://web.archive.org/web/20200815081900/http://www.linux-tutorial.info/modules.php?name=MContent&pageid=3}}
\footnote{\url{https://learn.microsoft.com/en-us/dotnet/framework/windows-services/introduction-to-windows-service-applications}}
\footnote{\url{https://developer.android.com/develop/background-work/background-tasks}}
. Background processes are a particularly underexplored contributor to software energy waste. These processes, such as autosaving, syncing, or logging, operate without direct user interaction, yet they generate consistent CPU, memory, and I/O activity over time. Although their energy footprint may appear minimal in isolation, their continuous nature and high execution frequency can lead to meaningful cumulative costs, especially at scale.

In this paper, we present a feature-level energy measurement process for background processes in interactive software. The process focuses on decomposing background features into their constituent operations (e.g., file writing, change tracking, metadata handling), and empirically benchmarking their energy consumption under controlled conditions. It is designed to support practical energy awareness during feature design, offering a lightweight and reproducible approach for profiling real-world implementations.

We instantiate the process via a case study of the autosave feature in three open-source Python-based desktop text editors. The study isolates key implementation differences across the editors, conducts 900 controlled measurements across multiple file sizes and autosave variants, and extracts practical insights on which design decisions drive energy usage. This case study was conducted originally in the context of the first author's MSc thesis at the University of Tartu, and serves here as a concrete instantiation of the presented process.

Our results show that save frequency, write buffering, and metadata operations significantly influence energy use in autosave feature, with one implementation reducing energy by up to 83\% through simple parameter adjustments. Based on these findings, we offer four actionable guidelines for developers implementing autosave or similar background features in Python.

The contributions of this paper include:
\begin{itemize}
  \item A reusable process that formalizes the steps needed to carry out a valid empirical experiment for measuring energy consumption of background processes.

  A fully published and reusable instantiation of the process as well
as its output\footnote{The MSc thesis by M. Küüsvek is available at \url{https://thesis.cs.ut.ee/250643ac-df5e-4993-9b73-a51d9b5d4d3f}}, including development, application, and evaluation of a measurement testbed for autosave energy consumption in real-world applications.
\item Actionable guidelines for developers to move towards greener implementations of the autosave feature.
\end{itemize}
These findings aim to inform greener software design practices and support energy-aware development of background features.

\section{Feature-level Energy Measurement Process}

Background processes such as logging, syncing, and silent updates are pervasive in modern software systems. Operating without direct user interaction, they are typically triggered by time intervals, system events, runtime conditions, or idle states. Despite their cumulative energy impact, these features are often overlooked during the design and evaluation phases. To address this gap, we propose a lightweight, feature-level energy measurement process for analyzing and benchmarking the energy behavior of background operations. The measurement process presented here builds upon established methodologies in software energy analysis \cite{hindle2015green,mancebo2021feetings,ardito2019methodological} but is tailored to the specific context of background processes. Whereas existing studies often evaluate energy consumption at the level of entire applications or functional modules, our process focuses narrowly on the isolated impact of background operations. This specialization enables a more precise examination of design trade-offs that might otherwise be obscured in broader analyses.

At its core, the process treats background features as composed of discrete atomic operations, the smallest functional units that can be individually isolated and profiled for energy consumption. It also emphasizes the importance of trigger mechanisms, which define the temporal or conditional context in which these operations are executed. By decomposing features into these elements, the process enables systematic energy profiling across implementations and contexts.

\begin{figure}[htbp]
\centering
\includegraphics[height=5.5cm,width=0.85\linewidth]{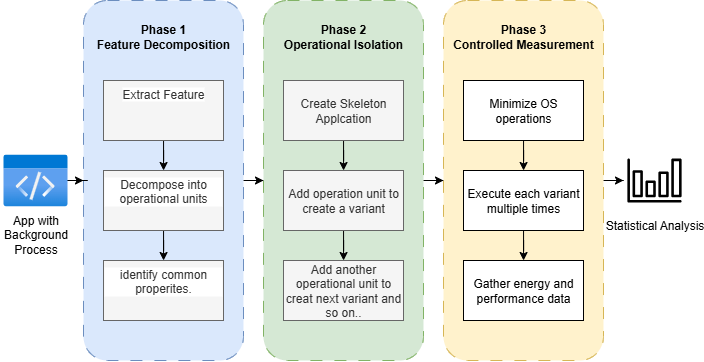}
\caption{Overview of Energy measurement process for Background Processes}
\label{fig:process}
\end{figure}

The feature-level energy measurement process proceeds in three Phases: 1) feature decomposition, 2) operational isolation, and 3) controlled measurement. A high-level schematic of the process is shown in Figure~\ref{fig:process}

\subsection{Phase 1: Feature Decomposition}
The first stage involves breaking down a background process into its constituent atomic operations, defined here as the smallest self-contained units of work that (a) perform a discrete functional task, and (b) can be instrumented and executed independently. Depending on the feature, these may include file access, state monitoring, data transformation, or network communication.

Alongside operational decomposition, common properties of the background process also need to be identified. These properties affect how demanding the background process is and how much energy it uses. The properties and their possible values are listed in Table~\ref{tab:bg-properties}. There is no consolidated list of common properties for background processes, and the ones shown in Table~\ref{tab:bg-properties} are extracted from industry\footnote{\url{https://dev.to/rajrathod/background-jobs-473j}\\ \url{https://learn.microsoft.com/en-us/azure/architecture/best-practices/background-jobs}\\ \url{https://developer.android.com/develop/background-work/background-tasks/persistent}} and online \cite{zsak2014impact,mercer1992introduction} sources. 

\begin{table}[ht]
\centering
\caption{Core Properties of Background Processes}
\label{tab:bg-properties}
\resizebox{.85\linewidth}{!}{%
\begin{tabular}{|l|p{9.5cm}|}
\hline
\textbf{Property} & \textbf{Description} \\
\hline
\textit{Trigger} & How the process is started: \textbf{Schedule-driven} (e.g., Time-based (i.e., periodic execution), Event-based (i.e., triggered after system or user input), Idle-based (i.e., syncing when the system is inactive), Reactive (e.g., triggering after buffer thresholds are exceeded). \\
\hline
\textit{Frequency} & How often the process runs: \textbf{Periodic} (fixed intervals), \textbf{Aperiodic} (irregular timing), or \textbf{Sporadic} (event-driven with minimum time between activations). \\
\hline
\textit{Persistence} & How long the process remains active or enabled: \textbf{Immediate} (short-lived), \textbf{Long-running}, \textbf{Deferrable} (can be delayed), or \textbf{Persistent} (remains across reboots). \\
\hline
\textit{Resource Usage} & The primary hardware resources consumed: e.g., \textbf{CPU}, \textbf{disk I/O}, \textbf{network}, \textbf{sensors}, etc. \\
\hline
\textit{Scope} & Whether the process executes \textbf{locally} (on-device operations) or interacts with \textbf{external systems} (e.g., network, cloud, peripherals). \\
\hline
\end{tabular}
}
\end{table}
%add footnote that these properties are extracted , and no proper classification exists.
\subsection{Phase 2: Operational Isolation}
The second stage focuses on isolating the core functionality of a background process within a skeleton application, a minimal working implementation that includes only the target feature, stripped of unrelated UI elements, services, or unrelated logic. This isolation allows for focused energy analysis by minimizing confounding variables.

Each atomic operation is introduced incrementally in a series of controlled application variants. For example, one variant might perform only buffer monitoring, while the next variant includes monitoring plus file I/O. This incremental setup supports analysis, enabling measurement of the marginal energy impact of each added component.

\subsection{Phase 3: Controlled Measurement}
The final stage involves collecting statistically reliable energy measurements in a consistent testing environment as described in existing literature \cite{hindle2015green,mancebo2021feetings,ardito2019methodological}.

\textit{Environment Setup and Preparation:}This includes minimizing OS and system background activity, fixing CPU frequency or disabling turbo modes (if applicable), and using high-resolution energy tools, such as Intel RAPL, via access layers like perf or pyRAPL. Such preparation reduces external variability and ensures that observed differences can be attributed to the feature under test.

\textit{Performing Measurements:} Each variant is executed multiple times across controlled parameters (e.g., data size, trigger frequency, I/O load). Multiple runs are required to achieve statistically reliable results and to smooth out noise introduced by the system environment. This process captures both the total energy consumed and the execution patterns of the background feature.

\textit{Data Analysis and Reporting:} Collected measurements are averaged and subjected to statistical treatment to account for natural variability across runs. Analysis focuses not only on total energy consumed but also on the relationship between operational design decisions (e.g., batching, buffering, trigger thresholds) and energy behavior. This allows for identifying trade-offs between performance and efficiency that might otherwise remain hidden.

\section{Case Study: Autosave Feature in Text-editors}

To evaluate the proposed feature-level energy measurement process, we applied it to the autosave feature in three real-world text editors: Mu\footnote{\url{https://github.com/mu-editor/mu}}, Leo\footnote{\url{https://github.com/leo-editor/leo-editor}}, and novelWriter\footnote{\url{https://github.com/vkbo/novelWriter}}. This case study serves as a practical demonstration of how the described process can be used to isolate and quantify energy usage in real software systems.
\subsection{Context}

Autosave is a widely used background feature in text editors, designed to protect users from data loss by saving content periodically or in response to user activity. While the functionality is similar across applications, the underlying implementation strategies vary significantly, particularly in terms of trigger mechanisms, file handling, and auxiliary operations such as logging and change detection.
% The primary objective of the case study is to apply the feature-level energy measurement process and determine how operational differences affect energy consumption, and to reveal actionable implementation tradeoffs.
Based on the context and objective following RQs are formed \\
\textbf{RQ1:} \textit{What operations within the autosaving feature are most energy-intensive?}\\
\textbf{RQ2:} \textit{How do different implementations of the autosave feature differ in energy consumption in selected text editors?}\\
\textbf{RQ3:} \textit{How do file size and save frequency impact the energy consumption of autosaving implementations in selected text editors?}

The case study focused exclusively on Python-based desktop text editors using the Qt\footnote{\url{https://doc.qt.io/qtforpython-6/}} framework to ensure consistency in architecture and UI behavior.

\subsection{Applying Phase 1: Feature Decomposition}

The autosave feature was decomposed into its core operations by inspecting the source code and runtime behavior of three open-source Python-based editors: Mu, Leo, and novelWriter. Three fundamental operations were consistently identified: 1) 
\textbf{file writing:} the act of persisting the active document or project state to disk, 2) \textbf{change detection:} logic to verify whether the document has changed since the last save, 3) \textbf{logging:} optional recording of autosave activity to console or log files. These operations were chosen for isolation because they represent the typical energy-consuming steps in autosave cycles. 
\begin{table}[htbp]
\centering
\caption{Autosave Implementation Differences}
\label{tab:autosave-impl}
\resizebox{.80\linewidth}{!}{%
\begin{tabular}{|p{2cm}|p{2.4cm}|p{2.4cm}|p{2.7cm}|p{2.3cm}|}
\hline
\textbf{Text Editor} & \textbf{Trigger Type} & \textbf{File Writing} & \textbf{Change Detection} & \textbf{Logging} \\
\hline
Mu & QTimer (5s) & Direct write + \texttt{fsync()} & Built-in Qt flag & File + Stream \\
\hline
novelWriter & QTimer (30s) & Temp file $\rightarrow$ overwrite & Custom flag & \texttt{stderr} only \\
\hline
Leo & Idle-time hook (300s) & Backup copy $\rightarrow$ overwrite & Manual flag & \texttt{print()} (optional) \\
\hline
\end{tabular}
}
\end{table}
Table~\ref{tab:autosave-impl} summarizes the most impactful implementation differences in selected editors
The autosave behavior across the three editors was governed by distinct trigger mechanisms. Each editor used a different type of trigger: Mu used a high-frequency timer (QTimer), triggering saves every 5 seconds. NovelWriter used a lower-frequency timer set to 30 seconds. Leo relied on an idle-time detection hook, firing autosave only after 5 minutes of user inactivity.
These differences affect not only when energy is consumed but also how frequently background processes interrupt system idle states. The case study captured these distinctions by configuring test variants that preserved original triggering behaviors wherever possible. Other common properties of the autosave feature across applications are as follows:

\begin{table}[ht]
\centering
\caption{Properties of Autosave Background Process}
\label{tab:bg-properties-autosave}
\resizebox{.70\linewidth}{!}{%
\begin{tabular}{|l|p{9.5cm}|}
\hline
\textbf{Property} & \textbf{Value} \\
\hline
\textit{Trigger} & Schedule driven \\
\hline
\textit{Frequency}& Periodic \\
\hline
\textit{Persistence} &Asynchronous \\
\hline
\textit{Resource} &usage Disk I/O \\
\hline
\textit{Scope}& Local \\
\hline
\end{tabular}
}
\end{table}
\subsection{Operational Isolation}
To isolate autosave energy use from unrelated system activity, each editor was reduced to a skeleton application. These lightweight Qt-based programs retained only the core autosave logic and a minimal text editing interface. All non-essential features, such as UI rendering, network access, or non-autosave I/O, were removed.

For each editor, three controlled variants were implemented using the skeleton application: 1) \textbf{Base:} Implementing only the file writing operation, 2) \textbf{Change:} Implementing file writing + change detection, 3) \textbf{Logging:} Implementing file writing + change detection + logging. These variants enabled incremental assessment of each operation’s energy cost while maintaining functional realism. Autosaving is carried out at identical intervals of 10 seconds (i.e., executed 12 times) in the skeleton application. To examine scalability, three file sizes were used in each test variant: 1) \textbf{0 KB:} Empty file (baseline behavior), 2) \textbf{5 KB:} Small document (code snippet or note), 3) \textbf{50 KB:} Large document (e.g., markdown or essay). All variants are available in the GitHub rep\footnote{\label{fn:repo}\url{https://github.com/MariaKuusvek/MSc-Thesis}}

\subsection{Controlled Measurement}
Each variant was tested in a controlled Linux environment using the Intel RAPL interface (via perf) to measure CPU and DRAM energy consumption. Autosave was triggered via each editor’s native mechanism, and artificial text edits were simulated to activate the feature reliably. Experimentation is carried out on a HP Elitebook with Lubuntu 24.04 LTS. Each measurement followed a fixed cycle: start the app, wait 5 sec, start Perf profiling, execute the test, let Perf finish, close the app, pause 5 min, then repeat.
Each configuration of test variant was executed 30 times, resulting in 900 total measurements ((3 editors × 3 variants × 3 file sizes × 30 repetitions) + 90 control test measurements).
\textbf{Comparative Profiling:} The data collected across all configurations was used to compare energy usage patterns across editors and variants. Statistical analysis was performed to reveal significant differences in energy consumption.

\begin{figure}[htbp]
\centering
\includegraphics[width=0.95\linewidth]{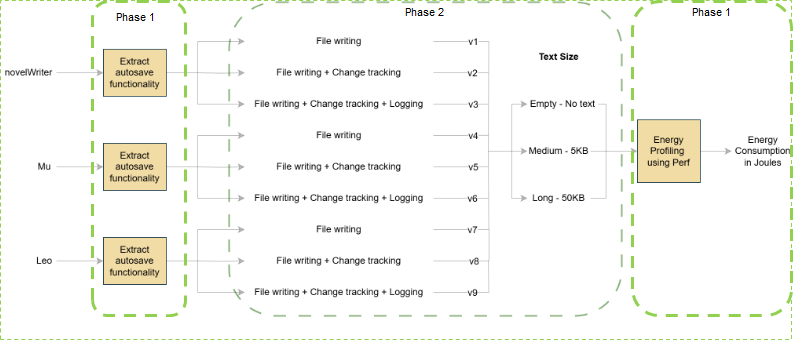}
\caption{Overview of energy measurement process in the context of case study}
\label{fig:Experiment}
\end{figure}

\section{Results}

We report on 900 measurements: 810 test case measurements and 90 control test measurements were carried out. The average values for each control test scenario were subtracted from the relevant main test to receive the delta energy value for autosaving. The full set of measurements taken is found in the CSV file in the GitHub\footref{fn:repo} repo. Below, only the conclusions are discussed\footnote{For full hypothesis testing results, see \url{https://docs.google.com/spreadsheets/d/14lV3lX50QL54Ocp-foSjeYGQ2VO_ROkp_NZSYdc1Z-w/edit?usp=sharing}}.

\textbf{Result of RQ1: \textit{(In-app Comparison)}} The basic file-writing operation was consistently the least energy-intensive across all editors, forming the baseline for autosave. Adding change tracking generally produced little to no additional cost, except in Mu, where medium and large files consumed an extra 5.83 J and 5.61 J compared to the base version, suggesting that its implementation of modified-flag checks is less efficient. Adding logging also had no statistically significant impact in any editor; in fact, raw means often indicated slight energy reductions, with the largest observed overhead being only 3.62 J, which is negligible compared to other operational differences.

\begin{figure}[htb]
\centering
\includegraphics[width=0.95\linewidth]{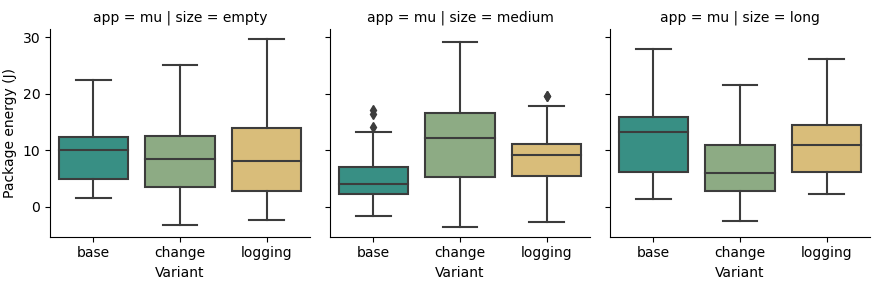}
\caption{Box plot of energy distribution in variants of Mu in all three file sizes.}
\label{fig:Fig8}
\end{figure}

\textbf{Result of RQ2: \textit{(Inter-app Comparison Results)} }The energy measurements revealed significant differences across autosave implementations, particularly in relation to trigger frequency and file write strategy. Mu, which uses high-frequency writes with fsync(), consistently consumed more energy than Leo and novelWriter. Leo's idle-triggered backup system was the most energy-efficient overall, especially for small files, while novelWriter performed well under moderate workloads.

\begin{table}[htb]
\centering
\caption{Summary of statistically significant pairwise differences in autosave energy consumption}
\resizebox{.85\linewidth}{!}{%
\begin{tabular}{lllll}
\hline
\textbf{Pairwise Comparison} & \textbf{File Size} & \textbf{p-value} & \textbf{Higher Energy User} & \textbf{$\Delta$ Energy (J)} \\
\hline
Mu base vs Mu change & 5 KB  & 0.0029   & Mu change            & 5.83  \\
Mu base vs novelWriter base & 5 KB & 0.003   & novelWriter base     & 6.84  \\
Mu base vs Leo base & 5 KB   & 0.046   & Leo base              & 4.04  \\
Mu change vs Leo change & 5 KB & 0.015   & Mu change            & 6.23  \\
Mu change vs novelWriter change & 5 KB & 0.000037 & Mu change     & 10.37 \\
Leo change vs Mu change & 50 KB & 0.0001 & Leo change           & 7.83  \\
Leo change vs novelWriter change & 50 KB & 0.016 & Leo change    & 5.01  \\
Leo logging vs novelWriter logging & 0 KB & 0.0155 & Leo logging & 5.12  \\
Mu logging vs novelWriter logging & 5 KB & 0.032 & Mu logging    & 3.75  \\
Mu logging vs Leo logging & 50 KB & 0.0021 & Mu logging         & 5.96  \\
\hline
\end{tabular}
}
\label{tab:pairwise}
\end{table}

\textbf{Result of RQ3: \textit{Scalability (File Size \& Save Frequency)}}

\begin{figure}[htb]
\centering
\includegraphics[width=0.95\linewidth]{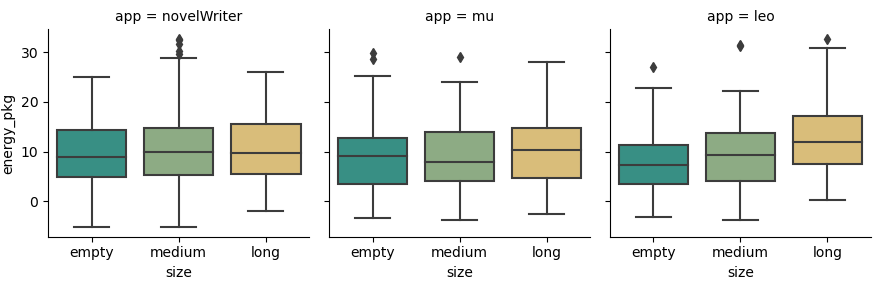}
\caption{Box plot of application-specific joule distribution by file size.}
\label{fig:Fig13}
\end{figure}

Scaling analysis showed that file size (see Figure~\ref{fig:Fig13}) had only a minor impact, while save frequency dominated energy consumption (see Table~\ref{tab:hourly}). All three editors consumed 0.8–0.9 J per save regardless of file size, but their default intervals produced stark contrasts: Mu’s 5-second interval scaled to 619 J/hour, far higher than novelWriter’s 100 J/hour at 30 seconds and Leo’s 11 J/hour at 5 minutes. Reducing Mu’s frequency to match novelWriter’s would cut its consumption by 83\%.

\begin{table}[ht]
\centering
\caption{Estimated hourly cost of autosaving in each application (RQ3).}
\label{tab:hourly}
\resizebox{.85\linewidth}{!}{%
\begin{tabular}{lccccc}
\toprule
\textbf{Editor (change+logging)} & \textbf{Avg J (12 saves)} & \textbf{Avg J per save} & \textbf{Frequency} & \textbf{Calls/hr} & \textbf{Joules/hr} \\
\midrule
novelWriter & 9.98 & 0.83 & 30 s & 120 & 99.6 \\
Mu          & 10.3 & 0.86 & 5 s  & 720 & 619.2 \\
Leo         & 11.07 & 0.92 & 300 s (idle) & 12 & 11.04 \\
\bottomrule
\end{tabular}
}
\end{table}

\section{Guidelines for Developers}

Based on the results of the case study, we propose four practical guidelines for developing an energy-efficient autosave feature in Python:
\begin{itemize}
    \item \textit{Minimize Save Frequency:} High-frequency autosave (e.g., every 5s) consumes significantly more energy than longer intervals. Use change-aware triggers or debounce logic to avoid redundant saves.
    \item \textit{Prefer in-place writes for autosave:}  If stronger crash safety is required, a temp-file→rename approach can be used, but it was still less efficient than in-place writes in our tests.
    \item \textit{Use dedicated change-tracking for autosave:} Avoid heavyweight/built-in modified flags; add a separate flag so autosave stops re-writing once changes cease, preventing unnecessary writes.
    \item \textit{Don’t fear logging:} Adding logging showed no statistically significant energy overhead across editors; treat logging placement as a maintainability/usability choice rather than an energy constraint.
\end{itemize}

\section{Threats to validity}
This process is intended to be flexible and extensible. While our use case focuses on a single background process in a very limited set of real-world applications, we believe the described process can be applied to other background features (such as syncing, telemetry, and auto-formatting etc.) and platforms, provided that atomic operations can be isolated or mocked, trigger conditions can be emulated, and energy usage can be measured meaningfully. However, there are limitations, such as the process assumes background features are discrete and isolatable. Highly integrated or event-cascading systems may require adaptation. Energy measurements are subject to system-level variability (e.g., caching, OS scheduling), even in controlled settings. Skeleton applications suggested in phase two of the process may not capture all side effects present in real-world systems. Despite these caveats, this process offers a practical pathway for incorporating energy-awareness into both research and design workflows.

Captured measurements provide clear comparative insights, but the study is limited to a single hardware configuration and focused on desktop applications written in Python and Qt. Results may vary on mobile devices, SSDs with different write performance, or platforms with advanced power management. Expanding the scope to other architectures or feature types is an important direction for future work.

\section{Related Work}

This section presents the most relevant literature related to the case study. Measuring the energy usage of software systems has gained momentum in sustainable computing research. Prior studies have explored energy implications from two perspectives: code-level implementation practices and system-level or environmental factors.

At the code level, Sahin et al. \cite{sahin2014code} and Şanlıalp et al. \cite{csanlialp2022energy} showed that even basic code refactorings (e.g., “extract local variable”, “simplify nested loops”) can influence energy usage by ±5–8\%. Chowdhury et al. \cite{chowdhury2018exploratory} isolated the impact of logging in Android apps, showing that high-frequency logging ($\geq10$ messages/sec) significantly increases energy usage due to disk I/O flushes, while low-frequency logging has a negligible impact. These findings support feature-level energy analysis as a method for guiding software design decisions.

Environment-level studies have examined broader structural attributes such as programming languages and software metrics. Mancebo et al. \cite{mancebo2021does} and Hindle \cite{hindle2015green} found that lines of code and churn metrics correlate with energy use. Pereira et al. \cite{pereira2021ranking} conducted a benchmark across 27 languages, finding that compiled languages like C and Rust are more energy-efficient than interpreted languages like Python. This contextualizes our case study, which focuses on implementations of background processes in Python, a widely used but energy-inefficient language.

Despite this progress, few studies have analyzed individual software features across multiple applications in a controlled, empirical setting. Jagroep et al. \cite{jagroep2016software} and Jimenez et al. \cite{jimenez2024analysing} looked at feature-level consumption, but did not isolate internal operations or background processes.

Autosave implementations differ in key dimensions: trigger mechanisms (timers vs. idle hooks), write strategies (buffered vs. unbuffered), and metadata handling. Yet no empirical study has systematically compared these factors in terms of their energy impact.

\section{Conclusion}

This paper introduced a reusable process for analyzing the energy behavior of background software features and demonstrated its application through a detailed case study of autosave implementations in three real-world text editors. By decomposing the feature, isolating operations, and profiling energy consumption under controlled conditions, we exposed measurable tradeoffs in trigger design, file handling, and auxiliary logic. Save frequency emerged as the dominant factor, with the Base variant in Mu consuming up to 83\% more energy than Leo under identical file sizes. Additional overhead from change detection and logging was observable but less pronounced; for example, logging added an average of 3–7\% energy cost depending on the editor and file size. The process proved effective for empirical energy evaluation and yielded four practical guidelines for designing greener autosave features. We believe this approach can be adapted to other background processes such as syncing, telemetry, and auto-formatting, and encourage further exploration in both research and development settings.

\begin{credits}
\subsubsection{\ackname} We would like to thank all internal reviewers at the University of Tartu for their valuable feedback.

\subsubsection{\discintname}
The authors have no competing interests to declare that
are relevant to the content of this article.
\end{credits}

%\bibliographystyle{splncs04}
%\bibliography{references}

\begin{thebibliography}{10}
\providecommand{\url}[1]{\texttt{#1}}
\providecommand{\urlprefix}{URL }
\providecommand{\doi}[1]{https://doi.org/#1}

\bibitem{andrae2017total}
Andrae, A.: Total consumer power consumption forecast. Nordic Digital Business Summit  \textbf{10}, ~69 (2017)

\bibitem{ardito2019methodological}
Ardito, L., Coppola, R., Morisio, M., Torchiano, M.: Methodological guidelines for measuring energy consumption of software applications. Scientific Programming  \textbf{2019}(1),  5284645 (2019)

\bibitem{chowdhury2018exploratory}
Chowdhury, S., Di~Nardo, S., Hindle, A., Jiang, Z.M.: An exploratory study on assessing the energy impact of logging on android applications. Empirical Software Engineering  \textbf{23}(3),  1422--1456 (2018)

\bibitem{hindle2015green}
Hindle, A.: Green mining: a methodology of relating software change and configuration to power consumption. Empirical Software Engineering  \textbf{20}(2),  374--409 (2015)

\bibitem{jagroep2016software}
Jagroep, E.A., van~der Werf, J.M., Brinkkemper, S., Procaccianti, G., Lago, P., Blom, L., van Vliet, R.: Software energy profiling: Comparing releases of a software product. In: Proceedings of the 38th International Conference on Software Engineering Companion. pp. 523--532 (2016)

\bibitem{jimenez2024analysing}
Jimenez, E., Pulido, C., Calero, C., Moraga, M.{\'A}., Garc{\'\i}a, F., Gordillo, A.: Analysing instagram's energy consumption: Tips for an eco-friendly use. IADIS International Journal on WWW/Internet  \textbf{22}(1) (2024)

\bibitem{jones2018stop}
Jones, N., et~al.: How to stop data centres from gobbling up the world’s electricity. nature  \textbf{561}(7722),  163--166 (2018)

\bibitem{mancebo2021does}
Mancebo, J., Calero, C., Garc{\'\i}a, F.: Does maintainability relate to the energy consumption of software? a case study. Software Quality Journal  \textbf{29}(1),  101--127 (2021)

\bibitem{mancebo2021feetings}
Mancebo, J., Calero, C., Garc{\'\i}a, F., Moraga, M.{\'A}., de~Guzm{\'a}n, I.G.R.: Feetings: framework for energy efficiency testing to improve environmental goal of the software. Sustainable Computing: Informatics and Systems  \textbf{30},  100558 (2021)

\bibitem{mercer1992introduction}
Mercer, C.W.: An introduction to real-time operating systems: Scheduling theory. Unpublished manuscript  (1992)

\bibitem{pang2015programmers}
Pang, C., Hindle, A., Adams, B., Hassan, A.E.: What do programmers know about software energy consumption? IEEE Software  \textbf{33}(3),  83--89 (2015)

\bibitem{pereira2021ranking}
Pereira, R., Couto, M., Ribeiro, F., Rua, R., Cunha, J., Fernandes, J.P., Saraiva, J.: Ranking programming languages by energy efficiency. Science of Computer Programming  \textbf{205},  102609 (2021)

\bibitem{sahin2014code}
Sahin, C., Pollock, L., Clause, J.: How do code refactorings affect energy usage? In: Proceedings of the 8th ACM/IEEE International Symposium on Empirical Software Engineering and Measurement. pp. 1--10 (2014)

\bibitem{csanlialp2022energy}
{\c{S}}anl{\i}alp, {\.I}., {\"O}zt{\"u}rk, M.M., Yi{\u{g}}it, T.: Energy efficiency analysis of code refactoring techniques for green and sustainable software in portable devices. Electronics  \textbf{11}(3), ~442 (2022)

\bibitem{zsak2014impact}
Zsak, N., Wolff, C.: Impact of video quality and wireless network interface on power consumption of mobile devices. arXiv preprint arXiv:1407.7667  (2014)

\end{thebibliography}
%

\end{document}